\documentclass[a4paper,fleqn]{cas-sc}

\usepackage{graphicx}
\usepackage{hyperref}
\usepackage[square, numbers, compress]{natbib}
\usepackage{amsmath}
\usepackage{amssymb}
\usepackage{bm} % for bold Greek letters 
\usepackage{xcolor}

\def\tsc#1{\csdef{#1}{\textsc{\lowercase{#1}}\xspace}}
\tsc{WGM}
\tsc{QE}
\tsc{EP}
\tsc{PMS}
\tsc{BEC}
\tsc{DE}
\renewcommand{\vec}[1]{{\mathbf #1}}

\begin{document}


\let\WriteBookmarks\relax
\def\floatpagepagefraction{1}
\def\textpagefraction{.001}
\shorttitle{Anisotropy of localized states in an anisotropic disordered medium}
\shortauthors{A. Goïcoechea, J. H. Page, and S.E. Skipetrov}
%\begin{frontmatter}

\title[mode = title]{Anisotropy of localized states in an anisotropic disordered medium}

\author[label1]{Antton Goïcoechea}
\fnmark[1]
\author[label1]{John H. Page}
\author[label2]{Sergey E. Skipetrov}

\address[label1]{Department of Physics and Astronomy, University of Manitoba, Winnipeg, Manitoba R3T 2N2, Canada}
\address[label2]{Univ. Grenoble Alpes, CNRS, LPMMC, 38000 Grenoble, France}

\fntext[fn1]{Present address: Institut Langevin, ESPCI Paris, CNRS UMR 7587, PSL University, 1 rue Jussieu, 75005 Paris, France}

\begin{abstract}
We study Anderson localization of a scalar wave in an ensemble of resonant point scatterers embedded in an anisotropic background medium. For uniaxial anisotropy of moderate strength, the mobility edges and the critical exponent of the localization transition are found to be unaffected by the anisotropy provided that the determinant of the anisotropy tensor is kept equal to one upon introducing the anisotropy. Localized modes have anisotropic spatial shapes although their anisotropy is weaker than the one expected from purely geometric considerations. The modes with the longest lifetimes are found to be the most anisotropic and their anisotropy increases with the size of the disordered medium.
\end{abstract}


\maketitle

\section{Introduction}
Anderson localization can be defined as the exponential suppression of transport of a wave due to strong disorder \citep{Anderson1958,abrahams201050,sheng2006introduction,akkermans2007mesoscopic}. Three-dimensional (3D) wave systems
%% obeying time-reversal symmetry and spin-rotation invariance
may exhibit a transition from the extended regime (diffusive transport) to the localization regime upon increasing the disorder at a constant frequency of the wave or upon varying the frequency at a fixed, sufficiently strong disorder \citep{Abrahams1979}. Several theoretical models exist to describe the localization transition, and to predict the evolution of the characteristic transport quantities as the transition point (also called ``mobility edge'') is approached (see, e.g., \cite{Evers2008,Wolfle2010} for reviews). Recently, a model based on the random Green's matrix, corresponding to a system of randomly distributed resonant point scatterers \citep{Rusek96,Goetschy2011,Goetschy2011EPL} has been extensively used. It allows different types of waves to be studied in a common framework, thus facilitating the identification of the key similarities and differences between light, sound and elastic waves. In particular, the model has provided a plausible explanation for the lack of experimental evidence of localization of light \citep{Skipetrov2014,SkipetrovPage2016} while confirming that Anderson localization of elastic waves is very much similar to that of scalar waves, as convincingly demonstrated in experiments \citep{Hu2008,Cobus2016,Cobus2018}. In addition, it has enabled the calculation of the critical exponent of the localization transition for scalar \citep{Skipetrov2016} and elastic \citep{SkipetrovBeltukov2018}  waves as well as for light scattered by atoms in a strong magnetic field \citep{Skipetrov2018PRL}.

In the present work, we adapt the theoretical framework based on the use of the random Green's matrix to study Anderson localization in anisotropic disordered media in which the speed of wave propagation depends on direction. Studies of Anderson localization in anisotropic media have long and rich history \citep{Wolfle1984, Bhatt1985, Li1989, Zhang1990, Abrikosov1994, Pana94, Milde1997, Milde2000, Kaas2008, Lopez2013, Pasek2017}. In particular, it was previously suggested that Anderson localization might be easier to achieve in anisotropic media because of the reduced effective dimensionality \citep{Abrikosov1994,Kaas2008}, whereas the scaling behavior is expected to remain independent of anisotropy \citep{Milde2000}. Additionally, the transport anisotropy (defined as the ratio of the diffusion tensor's components) has been predicted to be significantly reduced near the mobility edge based on the self-consistent theory of localization \citep{Piraud2014}, challenging earlier conclusions indicating that transport anisotropy is not affected by interference effects \citep{Wolfle1984,Bhatt1985}. Even though this prediction of significantly reduced transport anisotropy has recently been experimentally confirmed for ultrasonic waves propagating in anisotropic networks of aluminum beads \citep{Goicoechea2020}, the deficiency of the self-consistent theory to correctly describe the scaling properties of wave transport at the mobility edge and in the localized regime even in isotropic media \citep{Skipetrov2019Intensity} makes an alternative theoretical framework to study the problem of Anderson localization in anisotropic media highly valuable.

The point-scatterer model provides a solid foundation on which a description of general features of wave propagation in disordered media can be built without the need to rely on substantial approximations \citep{Lagendijk1996,VanTiggelen2020}. The idea is to consider multiple scattering of a wave in a large ensemble of randomly distributed resonant point scatterers. The theory is formulated in terms of the statistical properties of the Green's matrix, which depends on the relative distances between pairs of scatterers \citep{Rusek96,Skipetrov2014,Skipetrov2016}. Here, we focus on the model of scalar waves to study the effect of the anisotropy; some of the conclusions are expected to remain true for more complex vectorial elastic waves that have been used in recent experiments \citep{Goicoechea2020}. The anisotropy is introduced by considering an anisotropic Green's matrix, derived from the anisotropic wave equation. We show that within the accuracy of our calculations, a moderate anisotropy does not modify the locations of mobility edges. The critical exponent of the localization transition is found to be close to the value found for the isotropic case as well. The spatially localized eigenmodes have an anisotropic spatial distribution with an average anisotropy remaining below the ``geometric'' value that could be expected from the anisotropy of the medium. The spatial structure of the modes remains anisotropic at all length scales, although the anisotropy slightly decreases from localization centers towards the tails of the modes. The anisotropy is a monotonically increasing function of mode lifetime but exhibits a complicated dependence on the degree of mode localization measured by the inverse participation ratio of the mode. Our results demonstrate that the near isotropization of time-dependent (dynamic) wave transport near a mobility edge put forward by the previous theoretical \citep{Piraud2014} and experimental \citep{Goicoechea2020} works cannot be explained simply by the shape of individual localized modes, since the mode shapes themselves remain anisotropic.  

\section{Random Green's matrix model}
\label{model}

Consider an infinite anisotropic homogeneous medium in which $N \gg 1$ identical resonant point scatterers (resonance frequency $\omega_0$, resonance width $\Gamma_0 \ll \omega_0$) are embedded at random positions $\{ \mathbf{r}_n \}$ within a spherical volume $V$ of radius $R$ with a constant average number density $\rho = N/V$. We extend the previous work \citep{Rusek96,Skipetrov2014,Skipetrov2016} to study the multiple scattering of a monochromatic scalar wave in this system by considering the so-called Green's matrix $\mathbf{G}$ with elements
\begin{equation}
G_{nm} =  i \delta_{nm} + \left( 1 - \delta_{nm} \right) \frac{\exp{\left( i k_0 \sqrt{\mathbf{r}_{nm}\cdot \mathbf{A}^{-1}\cdot\mathbf{r}_{nm}} \right)}}{k_0 \sqrt{\mathbf{r}_{nm}\cdot \mathbf{A}^{-1}\cdot\mathbf{r}_{nm}}}
\label{G scal anis}
\end{equation}
The diagonal elements of this matrix describe the width of the resonance of a single scatterer (in units of $\Gamma_0/2$) while the off-diagonal elements are proportional to the solution of the  anisotropic wave equation \citep{Stevenson1990,Kaas2008}
\begin{equation}
\left[ \boldsymbol{\nabla} \cdot \mathbf{A} \cdot \boldsymbol{\nabla} + k_0^2 \right] \mathcal{G}(\mathbf{r}_n-\mathbf{r}_m) = \delta(\mathbf{r}_n-\mathbf{r}_m)
\label{weq}
\end{equation}
$G_{nm} = -(4\pi/k_0) \mathcal{G}(\mathbf{r}_n-\mathbf{r}_m)$ for $n \ne m$.
Here $\mathbf{A}$ is a $3 \times 3$ anisotropy tensor that we assume diagonal, $k_0=\omega_0 /c$ and $c$ are the wave number and the speed of the wave, respectively, in the homogeneous medium described by $\mathbf{A} = \mathbf{1}$, and $\mathbf{r}_{nm}= \mathbf{r}_{n}-\mathbf{r}_{m}$ is the distance between scatterers $n$ and $m$ located at $\mathbf{r}_{n}$ and $\mathbf{r}_{m}$, respectively.

It is obvious that Eqs.\ (\ref{G scal anis}) and (\ref{weq}) can be reduced to their isotropic versions by rescaling the spatial coordinates: $\mathbf{r} \to \mathbf{r}' = \sqrt{\mathbf{A}^{-1}} \cdot \mathbf{r}$. It may seem then that our anisotropic problem reduces to the isotopic one. However, this is not the case because the transformation of $\mathbf{r}$ to $\mathbf{r}'$ should also be applied to the scattering matrix $t(\omega)$ of an individual scatterer. Indeed, for a point scatterer that is supposed to model a spherically symmetric scattering object of small but non-zero size $a$, the scattering matrix is isotropic and is given by
\begin{equation}\label{tmatrix}
    t(\omega)= -\frac{4\pi}{k_0} \frac{\Gamma_0/2 }{\omega_0 - \omega - i\Gamma_0/2 }
 \end{equation}
with scalar $\omega_0 \sim c \sqrt{a/\alpha(0)}$ and $\Gamma_0 \sim \omega_0 k_0 a$, where $\alpha(0)$ is the static polarizability \citep{nieu92,DeVries98}. Transforming variables from $\mathbf{r}$ to $\mathbf{r}'$ deforms the spherically symmetric shape of the scatterer and makes the scattering matrix anisotropic. It is therefore possible to transform the problem of wave scattering by isotropic scatterers embedded in an anisotropic background medium to the problem of wave scattering by anisotropic scatterers in an isotropic background medium, but it is impossible to get rid of the anisotropy completely through such a transformation. The possibility of transferring the anisotropy from the background medium to the scatterers and back by a coordinate transformation allows us to link the present work to our recent experiments in which we have explored sound scattering in an elastic medium made of anisotropic, elongated scatterers with a direction-independent speed of wave propagation \citep{Goicoechea2020}.

If the diagonal elements of $\mathbf{A}$ are all equal ($A_{xx} = A_{yy} = A_{zz} = A$), the system becomes isotropic with a wave number $k_0'= k_0/\sqrt{A}$. This rescaling of $k_0$ can be absorbed in the normalization of the scatterer number density $\rho$ by $k_0'^3$ instead of $k_0^3$. To eliminate these trivial rescaling effects from consideration, we will restrict our analysis here to matrices $\mathbf{A}$ that have $\det \mathbf{A} = 1$.
For a uniaxial medium, we then have
\begin{eqnarray}
\mathbf{A} =
\begin{pmatrix}
\frac{1}{\sqrt{A_{zz}}} & 0 & 0 \cr
0 & \frac{1}{\sqrt{A_{zz}}} & 0 \cr
0 & 0 & A_{zz}
\end{pmatrix}
\label{amatric}
\end{eqnarray}
with a single parameter $A_{zz}$ measuring the anisotropy.

\section{Localization transitions}

To study localization transitions in the model introduced above, we will focus on the eigenvalues $\Lambda_{\alpha}$ and (right) eigenvectors $\bm{\psi}_{\alpha} = \{ \psi_{\alpha}^{(1)}, \psi_{\alpha}^{(2)}, \ldots, \psi_{\alpha}^{(N)} \}$ of the matrix $\mathbf{G}$. They obey
\begin{eqnarray}
\mathbf{G} \bm{\psi}_{\alpha} = \Lambda_{\alpha} \bm{\psi}_{\alpha},
\;\; \alpha = 1, \ldots, N
\label{eveq}
\end{eqnarray}
Real and imaginary parts of the eigenvalues $\Lambda_{\alpha}$ yield eigenfrequencies $\omega_{\alpha} = \omega_0 - (\Gamma_0/2) \mathrm{Re}\Lambda_{\alpha}$ and decay rates $\Gamma_{\alpha} = \Gamma_0 \mathrm{Im}\Lambda_{\alpha}$ of quasimodes, respectively. The degree of localization of eigenvectors can be quantified by their inverse participation ratios (IPR)
\begin{eqnarray}
\text{IPR}_{\alpha} =  \sum\limits_{m = 1}^{N} \left| \psi_{\alpha}^{(m)} \right|^4
\label{ipr}
\end{eqnarray}
where we assume that the eigenvectors are normalized:
\begin{eqnarray}
\sum\limits_{m = 1}^{N} \left| \psi_{\alpha}^{(m)} \right|^2 = 1
\label{norm}
\end{eqnarray}

\begin{figure}
\centering{
\includegraphics[width=0.47\columnwidth]{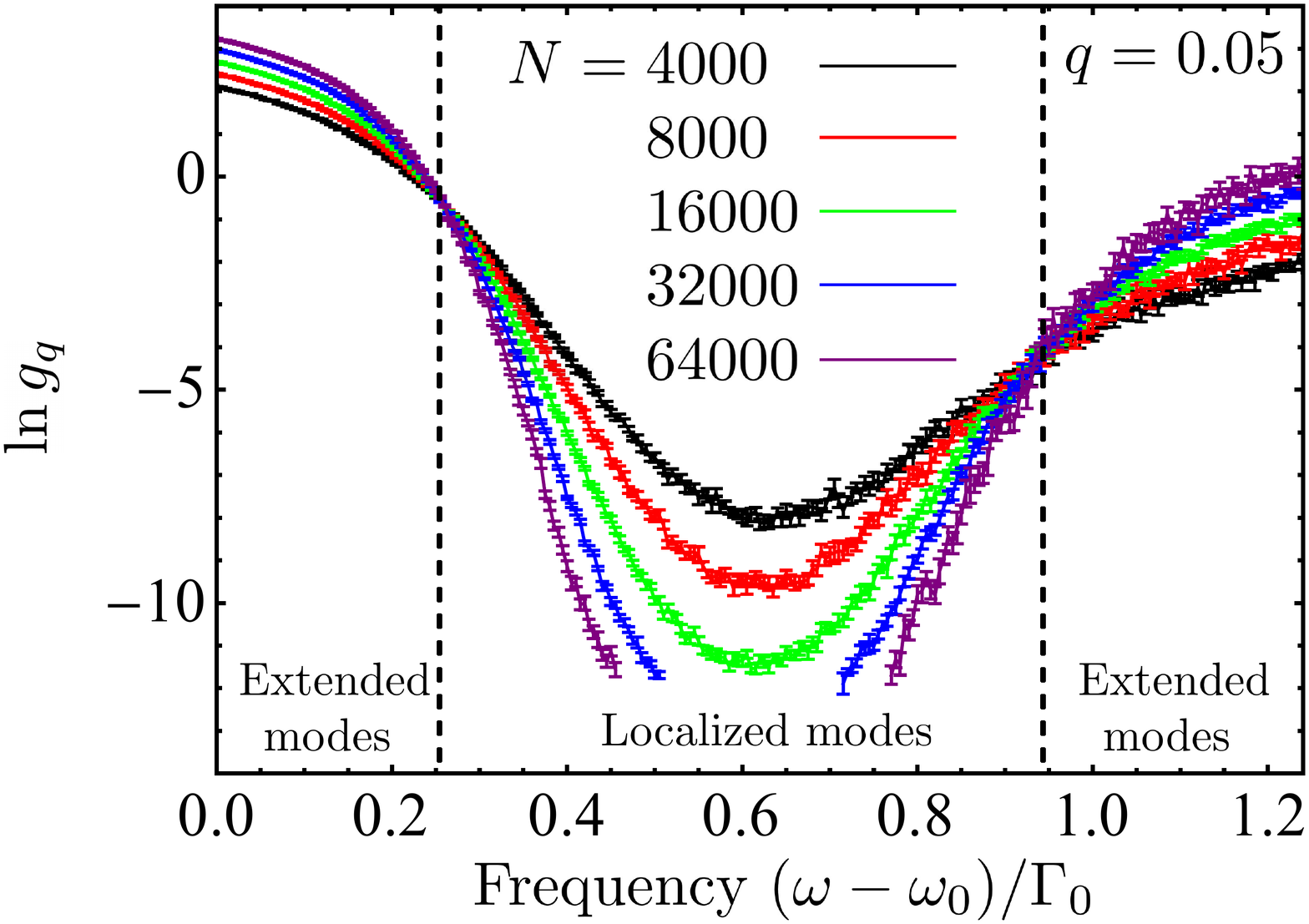}
}
%% \vspace{-3mm}
\caption{\label{fig_lngq}
Fifth percentile of the Thouless conductance as a function of frequency for different numbers of atoms $N$ at a fixed atomic number density $\rho/k_0^3 = 0.15$ and $A_{zz} = 2$. Averaging is performed over 800, 414, 414, 218 and 33 independent scatterer configurations for $N = 4000$, 8000, 16000, 32000, and 64000, 
respectively. Very low values of $\ln g_q \lesssim -12$ suffer from large statistical uncertainties and are not shown.  Dashed vertical lines at $(\omega-\omega_0)/\Gamma_0 = 0.254$ and 0.94 show positions of mobility edges determined by the finite-size scaling analysis illustrated in Fig.\ \ref{fig_me}.}
\end{figure}

Similarly to the isotropic case discussed in detail previously \citep{Skipetrov2016}, the eigenvectors $\bm{\psi}_{\alpha}$ of the matrix $\mathbf{G}$ (or ``quasimodes'', for short) may be extended over the whole system or localized in space. A transition between the two regimes is expected to take place at high scatterer densities $\rho$ and can be pinpointed by looking at the Thouless conductance associated with a quasimode $\bm{\psi}_{\alpha}$ \citep{Skipetrov2016}:
\begin{eqnarray}
g_{\alpha} = \frac{\Gamma_{\alpha}/2}{\langle |\omega_{\alpha} - \omega_{\alpha-1}| \rangle} =
\frac{\mathrm{Im}\Lambda_{\alpha}}{\langle |\mathrm{Re}\Lambda_{\alpha} - \mathrm{Re}\Lambda_{\alpha-1}| \rangle}
\label{gdef}
\end{eqnarray}
where the eigenvalues $\Lambda_{\alpha}$ are assumed to be ordered by their real parts. Obviously, $g_{\alpha}$ are random quantities and they should be characterized by a probability density $p_{\omega}(g)$ parametrized by the frequency $\omega$ of quasimodes. Because the distribution $p_{\omega}(g)$ can be quite wide, it is common to work with the distribution of the logarithm of conductance $p_{\omega}(\ln g)$. Instead of analyzing $p_{\omega}(\ln g)$, it is convenient to look at percentiles $\ln g_q$ defined by an integral relation
\begin{eqnarray}
q = \int\limits_{0}^{\ln g_q} p_{\omega}(\ln g) d(\ln g)
\label{perc}
\end{eqnarray}
The best known percentile is the fiftieth percentile ($q = 0.5$ or 50\%), known as the median. Analysis of the Anderson localization transition is more practical to perform by analyzing low-order percentiles ($q \ll 1$) that turn out to be less affected by irrelevant finite-size effects \citep{Skipetrov2016}. Figure \ref{fig_lngq} shows the fifth percentile ($q = 0.05$ or 5\%) obtained numerically by averaging over a large number of independent realizations of random matrices $\mathbf{G}$ of five different sizes $N$, all corresponding to the same number density of scatterers $\rho$. We observe that lines $\ln g_q(\omega)$ obtained for different $N$ all cross at two points. The appearance of such crossing points has been previously 
identified as a sign of mobility edges \citep{Skipetrov2016}. It turns out that frequencies $\omega_c$ at which these crossings take place are independent of $q$ for small $q$ (here we checked this property for $q \leq 0.05$). Using the definition (\ref{perc}) of $\ln g_q$, we conclude then that the small-$g$ wing of $p_{\omega}(\ln g)$ becomes independent of $N$ when $\omega = \omega_c$, which signals a localization transition.

\begin{figure}
\centering{
\includegraphics[width=0.47\columnwidth]{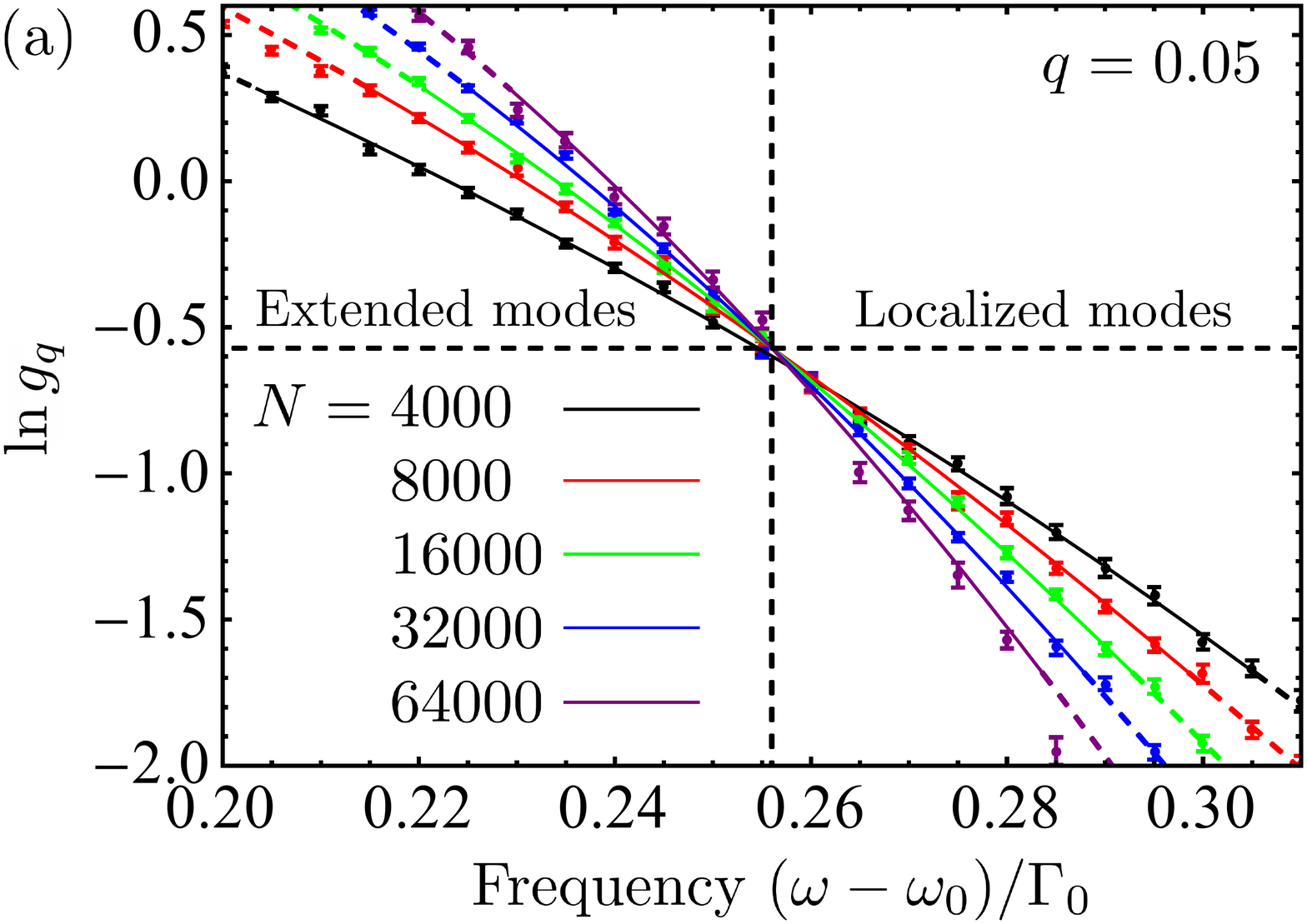}
\includegraphics[width=0.47\columnwidth]{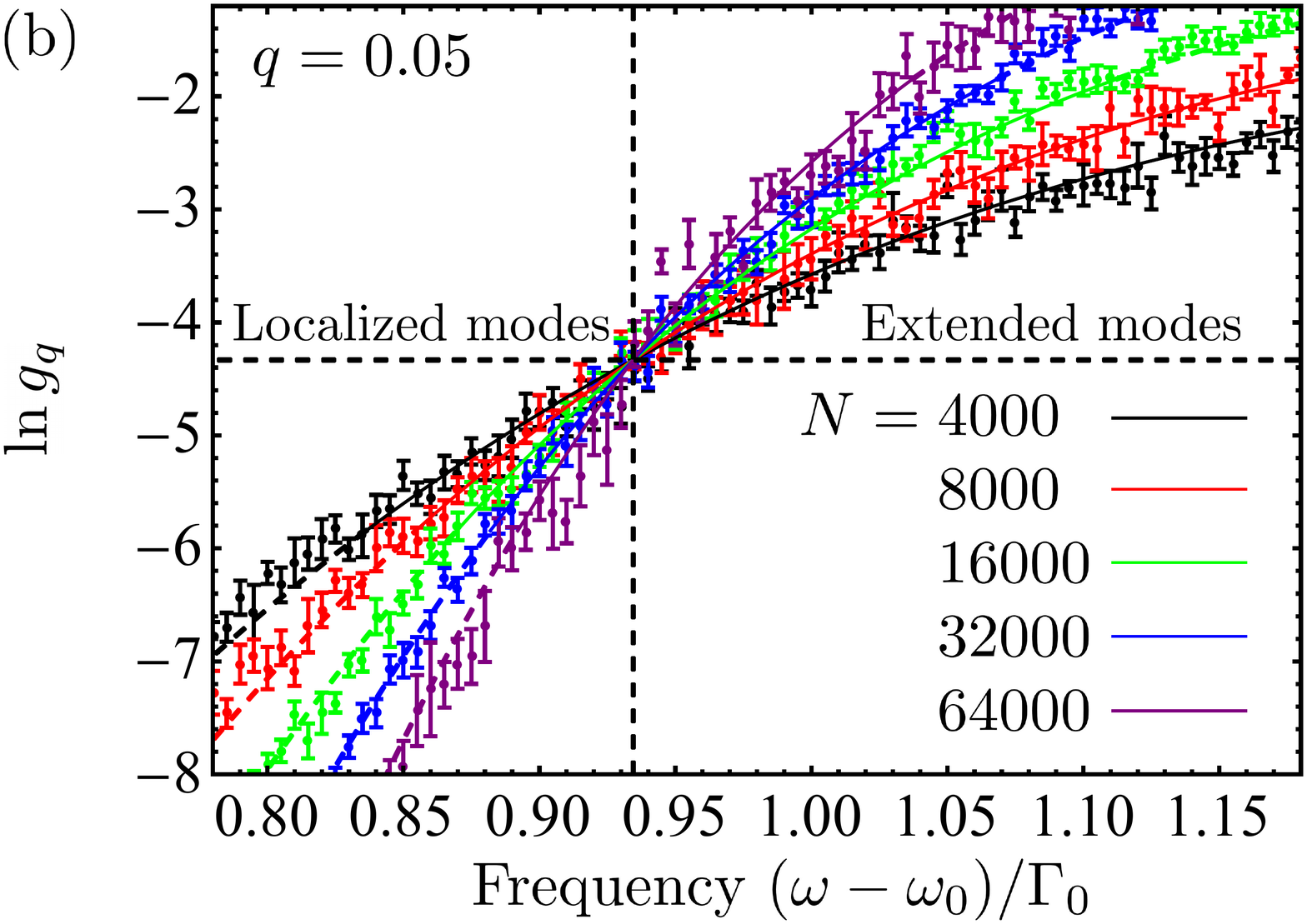}
}
%% \vspace{-3mm}
\caption{\label{fig_me}
Finite-size scaling analysis of the fifth percentile of the Thouless conductance at the low- (a) and high-frequency (b) mobility edges. Symbols with error bars show numerical data, lines are power-law fits performed following the procedure developed previously \citep{Skipetrov2016}. Dashed horizontal (vertical) lines show the critical conductances (mobility edges) deduced from the fits.}
\end{figure}

Analysis of $\ln g_q$ in the vicinity of crossing points using the standard finite-size scaling technique [see \citep{Skipetrov2016} for details] is illustrated in Fig.\ \ref{fig_me}. It involves assuming a single-parameter scaling and fitting the numerical data near the localization transition points $\omega_c$ to a power-law function of $R/\xi(\omega)$, where $\xi(\omega) \propto (\omega-\omega_c)^{-\nu}$ is the localization length and $\nu$ is the critical exponent. Similarly to the isotropic case, the numerical data near the low-frequency transition exhibit much weaker statistical fluctuations than the data near the high-frequency one. However, the positions of mobility edges $\omega_c$ can be reliably determined for both transitions albeit with different accuracies. The mobility edges are  shown in Figs.\ \ref{fig_lngq} and \ref{fig_me} by dashed vertical lines. Their positions averaged over $q \in [0.01, 0.05]$ are  $(\omega_c-\omega_0)/\Gamma_0 = 0.254 \pm 0.004$ and $(\omega_c-\omega_0)/\Gamma_0 = 0.94 \pm 0.02$, respectively. By comparing them with the values obtained earlier for an isotropic system ($\mathbf{A} = \mathbf{1}$) at the same density \citep{Skipetrov2016,Skipetrov2018} (\emph{e.g.}, in the more recent of these two papers \citep{Skipetrov2018}, the lower and upper mobility edges at this density were found to be $0.256 \pm 0.003$ and $0.935 \pm 0.011$, respectively), we conclude that, within the uncertainties of our analysis, the anisotropy does not modify the positions of mobility edges significantly, at least when the anisotropy is moderate ($A_{zz} \leq 2$).

Note that this result on the insensitivity of the mobility edges to anisotropy differs from some other conclusions in the literature \citep{Li1989, Zhang1990, Pana94, Milde1997, Milde2000, Lopez2013} that were based on calculations using different models, such as the Anderson tight-binding model, where the anisotropy that was introduced modifies many aspects of wave propagation.  Therefore, these other calculations do not reflect \emph{only} the effects of anisotropy \emph{per se}.  By contrast, by performing our analysis using a random Green’s matrix model with $\det \mathbf{A} = 1$, we are able to avoid such confounding effects such as, in our case, changes in the effective wave number, which would occur if this condition on $\mathbf{A}$ were not respected.  It is also worth noting that our finding (mobility edges which appear to be insensitive to anisotropy when $\det \mathbf{A} = 1$) is consistent with a prediction by Kaas et al. \citep{Kaas2008} for the Ioffe-Regel condition generalized to account for the presence of anisotropy. It is also consistent with recent numerical simulations for ultracold atoms that predict a universal scaling of the mobility edge that is independent of anisotropy when the disorder strength is appropriately rescaled \citep{Pasek2017}.   

The quality of our numerical data near the low-frequency transition [Fig.\ \ref{fig_me}(a)] is good enough to obtain a meaningful estimation of the critical exponent $\nu$ as well. The fits performed for $q = 0.01$--0.05 yield $\nu \simeq 1.6 \pm 0.2$, which is close to the value obtained for the same model in the isotropic case \citep{Skipetrov2016}. Thus, we do not observe any significant modification of the critical exponent due to the anisotropy of the disordered medium.

\section{Anisotropy of localized quasimodes}
\label{secparameter}

To characterize the anisotropy of the quasimodes $\bm{\psi}_{\alpha} = \{ \psi_{\alpha}^{(1)}, \psi_{\alpha}^{(2)}, \ldots, \psi_{\alpha}^{(N)} \}$, we compute central moments of order $2p$:
\begin{eqnarray}
\sigma_{\alpha x}^{(p)} &=& \left[ \sum\limits_{m = 1}^{N} \left| \psi_{\alpha}^{(m)} \right|^2
\left( x_m - X_{\alpha} \right)^{2p}
\right]^{1/2p}
\label{momentsx}
\\
\sigma_{\alpha y}^{(p)} &=& \left[ \sum\limits_{m = 1}^{N} \left| \psi_{\alpha}^{(m)} \right|^2
\left( y_m - Y_{\alpha} \right)^{2p}
\right]^{1/2p}
\label{momentsy}
\\
\sigma_{\alpha z}^{(p)} &=& \left[ \sum\limits_{m = 1}^{N} \left| \psi_{\alpha}^{(m)} \right|^2
\left( z_m - Z_{\alpha} \right)^{2p}
\right]^{1/2p}
\label{momentsz}
\end{eqnarray}
where
\begin{eqnarray}
\vec{R}_{\alpha} = \{X_{\alpha}, Y_{\alpha}, Z_{\alpha} \} = \sum\limits_{m = 1}^{N} \left| \psi_{\alpha}^{(m)} \right|^2
\vec{r}_m
\label{center}
\end{eqnarray}
is the localization center of the quasimode $\bm{\psi}_{\alpha}$.

In the anisotropic medium defined by Eq.\ (\ref{amatric}), the anisotropy of the shape of a given quasimode $\bm{\psi}_{\alpha}$ can be characterized by a parameter
\begin{eqnarray}
{\cal A}_{p} = \frac{1}{2} \left[ \frac{\sigma_{\alpha z}^{(p)} - \sigma_{\alpha x}^{(p)}}{\frac12(\sigma_{\alpha z}^{(p)} + \sigma_{\alpha x}^{(p)})}
+ 
\frac{\sigma_{\alpha z}^{(p)} - \sigma_{\alpha y}^{(p)}}{\frac12(\sigma_{\alpha z}^{(p)} + \sigma_{\alpha y}^{(p)})}
\right]
\label{ratio}
\end{eqnarray}

Obviously, being defined for a given quasimode, ${\cal A}_{p}$ is not strictly equal to zero even in an isotropic medium. However, it should vanish on average if $A_{zz} = 1$. One can also check that for both anisotropic Gaussian
\begin{eqnarray}
\psi(\vec{r}) =  \frac{1}{\sqrt{\pi^{3/2} \sigma_x^2 \sigma_z}}
\exp\left( -\frac{x^2}{2 \sigma_x^2} - \frac{y^2}{2 \sigma_x^2} - \frac{z^2}{2 \sigma_z^2} \right)
\label{gauss}
\end{eqnarray}
and exponential
\begin{eqnarray}
\psi(\vec{r}) =  \frac{1}{\sqrt{8 \sigma_x^2 \sigma_z}}
\exp\left( -\frac{|x|}{2 \sigma_x} - \frac{|y|}{2 \sigma_x} - \frac{|z|}{2 \sigma_z}\right)
\label{exp}
\end{eqnarray}
functions, we obtain
\begin{eqnarray}
{\cal A}_p = \frac{\sigma_z - \sigma_x}{\frac12(\sigma_z + \sigma_x)}
\label{ratiomodels}
\end{eqnarray}
independent of $p$. This shows that ${\cal A}_p$ is a good candidate for properly capturing the anisotropy for various quasimode shapes.

It follows from the definition (\ref{ratio}) that ${\cal A}_p$ becomes more sensitive to tails of eigenfunctions for increasing $p$. Typically, the largest contribution to $\sigma_{\alpha \mu}^{2p}$ (with $\mu = x$, $y$ or $z$) comes from $\mu \sim \sqrt{p} \sigma_{\mu}$ and $\mu \sim 2 p \sigma_{\mu}$ for Gaussian and exponential quasimode profiles, respectively. Thus, analyzing ${\cal A}_p$ for different values of $p$ allows anisotropy at different length scales to be probed.

\begin{figure*}
\includegraphics[width=0.97\textwidth]{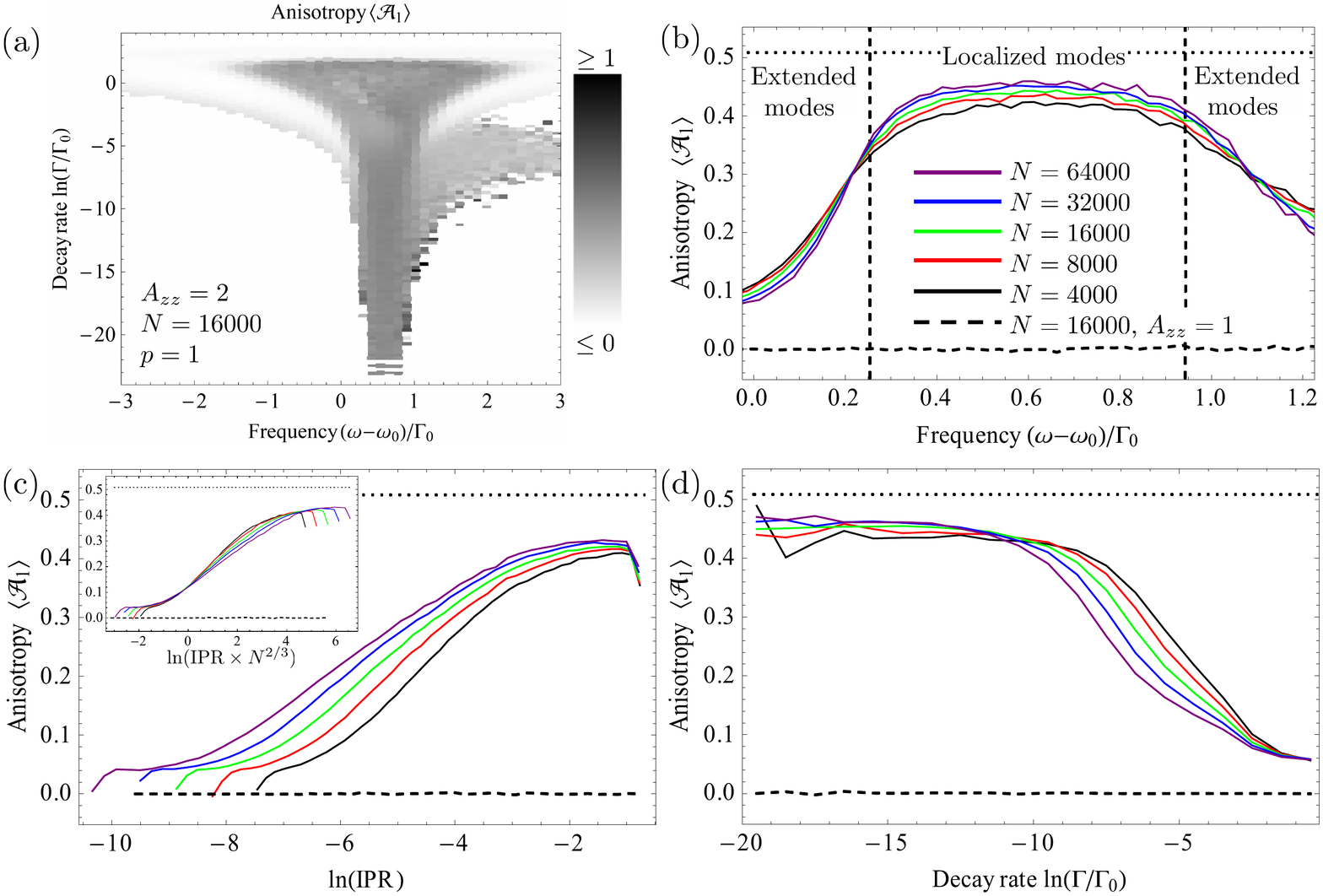}
%% \vspace{-5mm}
\caption{\label{fig_p=1}
Average anisotropy parameter $\langle {\cal A}_p \rangle$ for $\rho/k_0^3 = 0.15$, $A_{zz} = 2$ and $p = 1$. (a) Grayscale plot of $\langle {\cal A}_1 \rangle$ for $N = 16000$ scatterers in a sphere.  The white color corresponds mainly to regions without eigenvalues; rare spots with negative $\langle {\cal A}_1 \rangle$ appear due to insufficient averaging. 
(b--d) $\langle {\cal A}_1 \rangle$ as a function of frequency $(\omega-\omega_0)/\Gamma_0$ (b), $\text{IPR}$ (c) and decay rate $\ln(\Gamma/\Gamma_0)$ (d). Lines of different colors correspond to different numbers of scatterers [see the legend in (b)]. Horizontal dashed lines $\langle {\cal A}_1 \rangle \simeq 0$ show the result obtained for the isotropic case $A_{zz} = 1$ and $N = 16000$. Dotted horizontal lines show the anisotropy expected from geometrical considerations: $\langle {\cal A}_1 \rangle = 2(\sqrt{A_{zz}}-\sqrt{A_{xx}})/(\sqrt{A_{zz}}+\sqrt{A_{xx}}) \simeq 0.51$. Vertical dashed lines in (b) show the mobility edges. Inset of (c) illustrates scaling with $N$. Averaging is performed over the same numbers of independent scatterer configurations as in Fig.\ \ref{fig_lngq} and over 573 configurations for $N = 16000$ in the isotropic case.}
\end{figure*}

\begin{figure*}
\includegraphics[width=0.97\textwidth]{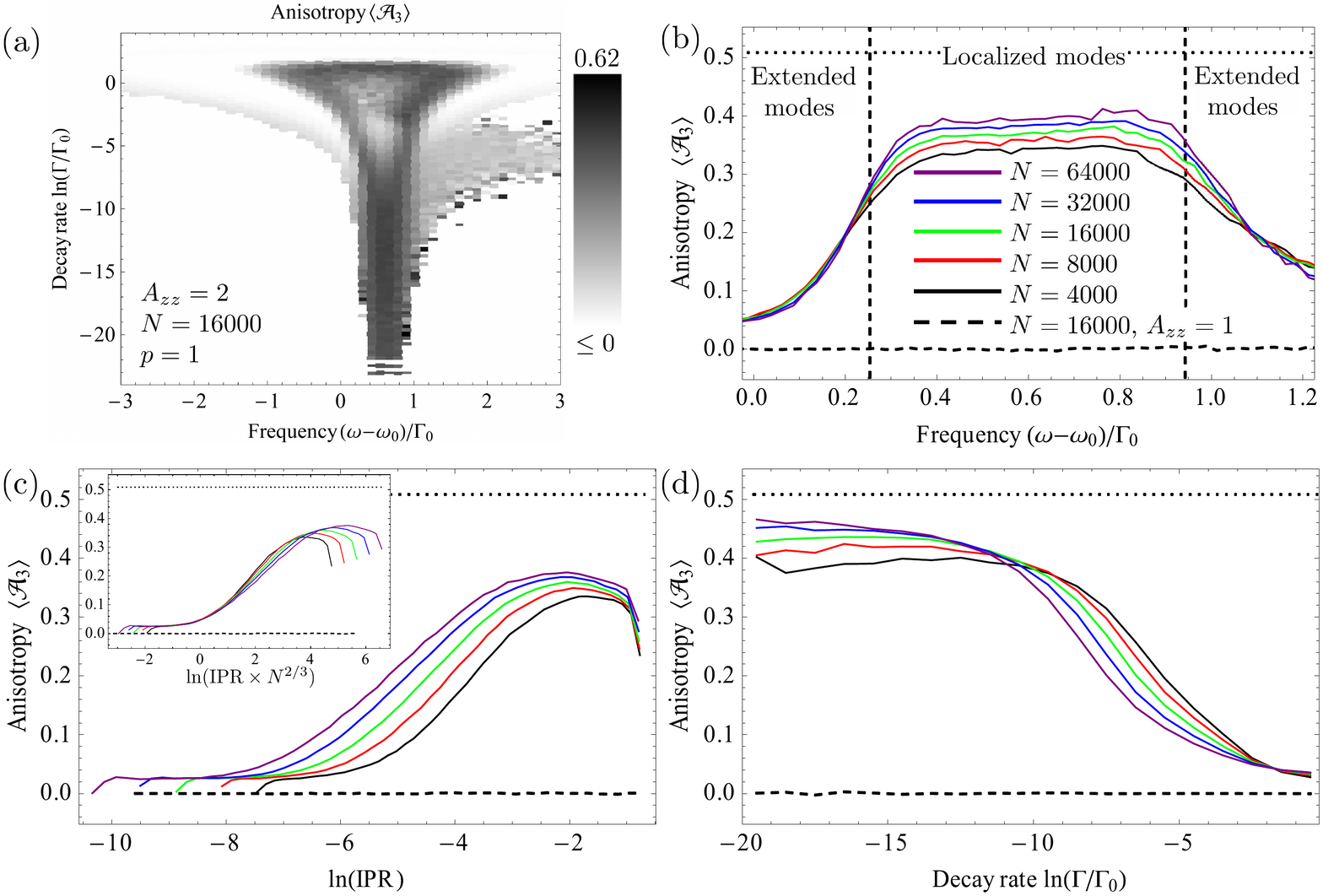}
%% \vspace{-5mm}
\caption{\label{fig_p=3}
Same as Fig.\ \ref{fig_p=1} but for $p = 3$.}
\end{figure*}

We compute the anisotropy parameter ${\cal A}_p$ for the same scatterer number density and the same numbers of scatterers as in Figs.\ \ref{fig_lngq} and \ref{fig_me} and average it over many independent random scatterer configurations. The results are shown in Figs.\ \ref{fig_p=1} and \ref{fig_p=3} for $p = 1$ and $p = 3$, respectively. They are compared to the isotropic case ($\mathbf{A} = \mathbf{1}$) for which we obtained dashed lines $\langle {\cal A}_p \rangle \simeq  0$, as well as with a value that could be expected from purely geometric considerations by putting $\sigma_{\alpha z}^{(p)} = \sqrt{A_{zz}}$ and $\sigma_{\alpha x}^{(p)} = \sigma_{\alpha y}^{(p)} = \sqrt{A_{xx}} = \sqrt{A_{yy}}$ (dotted lines). This geometric value of ${\cal A}_p$, equal to 0.51 in the considered case of $A_{zz} = 2$, corresponds to a situation in which the modes of the isotropic system are ``pinned'' to the scatterers while the ensemble of the latter is being stretched-shrunk to perform a coordinate transformation  $\mathbf{r} \to \mathbf{r}' = \sqrt{\mathbf{A}^{-1}} \cdot \mathbf{r}$. As discussed in Sec.\ \ref{model}, such a coordinate transformation is not sufficient to describe scattering in an anisotropic medium because it does not account for the transformation of the scattering matrix of individual scatterers. However, it provides a reasonable reference value of ${\cal A}_p$ for localized modes, as we will see later.    

We first discuss $\langle {\cal A}_p \rangle$ as a function of frequency and decay rate, as shown in Figs.\ \ref{fig_p=1}(a) and \ref{fig_p=3}(a). Note that  ${\cal A}_p$ is a strongly fluctuating quantity and can take negative or large positive values even in the considered case of $A_{zz} = 2$. On average, we find $0 < \langle {\cal A}_p \rangle < 1$ but there are not always enough eigenvectors for the averaging to converge for all points in Figs.\ \ref{fig_p=1}(a) and \ref{fig_p=3}(a). Thus, we restrict the gray scale of these figures to the ranges $[0, 1]$ and $[0, 0.62]$, respectively, by showing the values below 0 (above 1) by the same gray scale as 0 (1). Indeed, rare negative values of $\langle {\cal A}_p \rangle$ and $\langle {\cal A}_1 \rangle > 1$ result from insufficient averaging and are of no interest for us here. A bright spot in the middle of  \ref{fig_p=3}(a) shows that $\langle {\cal A}_3 \rangle$ drops for quasimodes within the band of localized states $(\omega_c-\omega_0)/\Gamma_0 \in [0.254, 0.94]$ and having moderate decay rates. No such a spot is visible in Fig.\ \ref{fig_p=1}(a), suggesting a significant dependence on $p$. However, the drop of $\langle {\cal A}_3 \rangle$ is not visible when we project the data on the real axis and plot $\langle {\cal A}_3 \rangle$ as a function of frequency [Fig.\ \ref{fig_p=3}(b)]. For both $p = 1$ and $p = 3$, we observe that $\langle {\cal A}_p \rangle$ is almost constant within the band of localized states and drops to zero outside of this band. This isotropization of extended modes is explained by the fact that they occupy the whole available space and thus extend equally in all directions: $\sigma_{\alpha \mu} \sim R$ for any $\mu = x$, $y$, $z$, independent of the anisotropy of the scattering medium. In contrast to extended modes, localized modes show a considerable degree of anisotropy as witnessed by $\langle {\cal A}_p \rangle$ growing near mobility edges and reaching a maximum inside the band of localized states. However, even for localized quasimodes the anisotropy remains weaker than the one that can be expected from purely geometric considerations ($\langle {\cal A}_p \rangle \simeq 0.51$ shown by dotted horizontal lines). $\langle {\cal A}_p \rangle$ increases with $N$ and thus it cannot be excluded that the geometric limit may be reached for $N \to \infty$. However, the increase is very slow and there is little hope that the limiting value of $\langle {\cal A}_p \rangle$ could be attained for numerically manageable values of $N$. We also notice that  $\langle {\cal A}_3 \rangle$ is slightly less than $\langle {\cal A}_1 \rangle$, indicating that the mode profile tends to become less anisotropic as the distance from the mode localization center increases. 

To further explore the relation between localization and anisotropy of quasimodes, we plot $\langle {\cal A}_p \rangle$ as a function of mode IPR. First, the comparison of Figs.\ \ref{fig_p=1}(c) and \ref{fig_p=3}(c) confirms the decrease of $\langle {\cal A}_p \rangle$ with $p$ and thus the less anisotropic mode structure at larger distance from the localization center. Second, $\langle {\cal A}_p \rangle$ reaches its largest values for $\text{IPR} \sim 0.1$--0.4 whereas it tends to slightly decrease for states with the maximal $\text{IPR}$. This suggests that modes localized on clusters of a few scatterers and, in particular, on pairs of closely located scatterers ($\text{IPR} = 0.5$ for such modes) exhibit less anisotropy, with this effect being much more pronounced for $\langle {\cal A}_3 \rangle$ than for $\langle {\cal A}_1 \rangle$. Similarly to Figs.\ \ref{fig_p=1}(b) and \ref{fig_p=3}(b), $\langle {\cal A}_p \rangle$ increases with the number of scatterers $N$ (and hence with the sample size $R$) indicating that it may tend to its geometric limit for an infinite medium.

The dependence of $\langle {\cal A}_p \rangle$ on IPR exhibits an intriguing plateau, $\langle {\cal A}_p \rangle \simeq 0.03$--0.04, at low IPR values. The plateau is independent of $N$ and extends from $\text{IPR} \sim 1/N$ to $\text{IPR} \sim 1/N^{2/3}$ as can be demonstrated by replotting $\langle {\cal A}_p \rangle$ as a function of $\text{IPR} \times N^{2/3}$, see the insets of Figs.\ \ref{fig_p=1}(c) and \ref{fig_p=3}(c). Such a scaling with $N$ suggests that the existence of the plateau is related to surface modes of which the number scales as $N^{2/3}$. Roughly speaking, $1/N < \text{IPR} < 1/N^{2/3}$ includes weakly localized modes in the bulk and extended modes at the surface of the medium, whereas $\text{IPR} > 1/N^{2/3}$ corresponds exclusively to localized states, both in the bulk and on the surface. It is, however, unclear why the interplay between bulk and surface modes gives rise to a well-defined, $N$-independent plateau of $\langle {\cal A}_p \rangle$. 

For completeness, we also plot $\langle {\cal A}_p \rangle$ as a function of decay rate $\Gamma$ of the modes [Figs.\ \ref{fig_p=1}(d) and \ref{fig_p=3}(d)].  Interestingly enough, here $\langle {\cal A}_p \rangle$ increases with $N$ for $\ln(\Gamma/\Gamma_0) \lesssim -11$ whereas it decreases with $N$ for larger $\ln(\Gamma/\Gamma_0)$ and then shows yet another change of behavior for $\ln(\Gamma/\Gamma_0)$ between $-2$ and 0.  Despite this quite complicated dependence on $N$, these figures show clearly that the longest-lived modes have the largest anisotropy, although, because of the subtle connection between mode lifetime and IPR, one should not na\"ively assume that the longest-lived modes correspond to all of the most localized ones.    

\section{Relation to transport properties}

Our results indicate that localized modes have anisotropic spatial structure, even though their anisotropy remains weaker than one could expect from purely geometric considerations. At the same time, previous calculations \citep{Piraud2014} and measurements \citep{Goicoechea2020} of wave propagation in anisotropic disorder find a tendency towards isotropization of transport properties when the mobility edge is approached from the diffusion side, thus suggesting that localized states might also be nearly isotropic. To understand the relation between these two results, we note that in a transport experiment, one is interested in the propagation of an initial excitation through the disordered medium. For our model of point scatterers, the vector $\vec{u}(t) = \{ u^{(1)}(t), u^{(2)}(t), \ldots, u^{(N)}(t)  \}$ of wave amplitudes on the scatterers can be expanded over the $N$ eigenvectors $\bm{\psi}_{\alpha}$ of the matrix $\mathbf{G}$:
\begin{eqnarray}
\vec{u}(t) = 
\sum\limits_{\alpha = 1}^N
C_{\alpha}(t) \bm{\psi}_{\alpha}
\label{dynamics}
\end{eqnarray}
where the magnitudes and the time dependences of $C_{\alpha}(t)$ are determined by the overlap of the initial excitation with the eigenvectors and its frequency spectrum. For a spherical ensemble of scatterers considered in this work and the initial excitation (source) at its center, the expansion of the wave with time can be characterized by parameters
\begin{eqnarray}
\eta_{x}^2(t) = \sum\limits_{m = 1}^{N} \left| u^{(m)}(t) \right|^2
x_m^2,\;\;\;
\eta_{y}^2(t) = \sum\limits_{m = 1}^{N} \left| u^{(m)}(t) \right|^2
y_m^2,\;\;\;
\eta_{z}^2(t) = \sum\limits_{m = 1}^{N} \left| u^{(m)}(t) \right|^2
z_m^2
\label{expansion}
\end{eqnarray}
that we intentionally define by analogy with Eqs.\ (\ref{momentsx}--\ref{momentsz}) for $p = 1$. Substituting Eq.\ (\ref{dynamics}) into Eqs.\ (\ref{expansion}), the latter become 
\begin{eqnarray}
\eta_{x}^2(t) = \sum\limits_{\alpha, \beta = 1}^N
C_{\alpha}^*(t)
C_{\beta}(t)
\sum\limits_{m = 1}^{N}
\psi_{\alpha}^{(m)*}
x_m^2
\psi_{\beta}^{(m)}
\label{expansion2}
\end{eqnarray}
and similarly for $\eta_{y}^2(t)$ and $\eta_{z}^2(t)$.

Averaging of Eq.\ (\ref{expansion2}) over different scatterer configurations is not an easy task. Even if we decouple the average of $C_{\alpha}^*(t)
C_{\beta}(t)$ from the rest of the righthand side of Eq.\ (\ref{expansion2}), cross-terms $\langle \psi_{\alpha}^{(m)*}
x_m^2 \psi_{\beta}^{(m)} \rangle$ still remain.
Their contribution to $\eta_{x}^2(t)$ is negligible only in the diagonal approximation that becomes valid in the limit of a very long time $t$ exceeding the typical time of wave propagation through the medium $t_D \sim R^2/D$ (with some residual diffusion coefficient $D$) and the Heisenberg time $t_H \sim \text{DOS}(\omega) V$ (with $\text{DOS}(\omega)$ the density of states at frequency $\omega$). For $t \gg t_D, t_H$, Eq.\ (\ref{dynamics}) is dominated by a single, longest-lived eigenvector and the anisotropy of the wave intensity distribution coincides with the anisotropy of this eigenvector. However, such a long-time regime is difficult to reach in an experiment because measured signals decay in time, with the inevitable presence of absorption in experiments making access to this very long-time regime all the more problematic \cite{Goicoechea2020}.  The long-time limit $t \gg t_D, t_H$ is also irrelevant for a wavepacket expansion in an unbounded system where both $t_D$ and $t_H$ are infinite \cite{Piraud2014}. Thus the analysis of the present work is insufficient to enable relevant conclusions to be made about  transport properties. The works \citep{Piraud2014,Goicoechea2020} call for additional analysis.     

\section{Conclusions}
\label{secresults}

In the present paper, we use the resonant point scatterer model to study Anderson localization of a scalar wave in a 3D anisotropic disordered medium where the velocity of wave propagation depends on direction. To avoid trivial effects due to rescaling of the effective orientation-averaged velocity, we restrict our consideration to the anisotropy tensor $\mathbf{A}$ obeying $\det\mathbf{A} = 1$. Guided by the experimentally realizable conditions \citep{Goicoechea2020}, we only study uniaxial media with moderate anisotropy.

Under the conditions described above, we find that the Anderson localization transition takes place in very much the same way as in the equivalent isotropic medium. In particular, the mobility edges do not shift and the critical exponent remains unchanged, at least within the accuracy of our analysis. This is in agreement with previous analytic \citep{Kaas2008} and numerical \citep{Milde2000} results, although we cannot exclude that the situation may change for stronger anisotropy when the medium becomes closer to an effective 1D (for $A_{zz} \gg 1$) or 2D (for $A_{zz} \ll 1$) system. To further investigate the effects of anisotropy, we have chosen to study one of the most basic properties, namely the nature of the spatial structure of localized modes.  We find that localized modes exhibit considerable anisotropy, and have slightly more anisotropy near their localization centers than in their tails. The modes with longest life times are found to be the most anisotropic whereas the relation between the anisotropy and the degree of localization is less univocal: the anisotropy is maximal for modes with $\text{IPR} \sim 0.1$--0.4 but decreases for the most localized states with  $\text{IPR} = 0.5$. An interplay between surface and bulk modes gives rise to a plateau of the anisotropy parameter $\langle {\cal A}_p \rangle \simeq 0.03$--0.04 for modes with IPR between $1/N$ and $1/N^{2/3}$. In all cases, the anisotropy of mode shapes remains smaller than the geometric one calculated directly from the anisotropy of the disordered medium, even though a tendency for the localized mode anisotropy to increase with the medium size is found.   

The possibility of approximately mapping the anisotropy in wave propagation velocity to the geometric anisotropy of individual scattering units allows us to compare our results with those of the recent experiment performed with elastic waves scattered in a strongly heterogeneous, anisotropic medium \citep{Goicoechea2020}. However, it turns out that the properties of individual modes analyzed in the present paper are not sufficient to establish a direct link between the two. The challenges in making this link are perhaps best illustrated by the case of extended states in the diffuse regime of an anisotropic material, for which  the modes are isotropic, as we have shown, but the transport is not.  More work is needed to understand if the point-scatterer model can be applied to model the general features of the experiment and interpret its results.  

Finally, the question of universality of our conclusions remains largely open. Indeed, the anisotropy can be introduced in a disordered medium in many different ways and it is not at all obvious whether its impact on Anderson localization should be the same in all cases. Even for the point-scatterer model used in this work, many interesting questions require further investigation. It would be, for example, interesting to study the evolution of the localization--delocalization phase diagram of the model as a function of the anisotropy tensor $\mathbf{A}$ and extend the analysis to stronger, biaxial, or non-diagonal anisotropies.

\section*{Acknowledgements}

Support from the Natural Sciences and Engineering Research Council of Canada’s Discovery Grant Program (RGPIN-2016-06042) is gratefully acknowledged.

\bibliographystyle{elsarticle-num}

\bibliography{references_clean.bib}

\end{document}